# A Value-Centered Exploration of Data Privacy and Personalized Privacy Assistants

**Abstract:** In the current post-GDPR landscape, privacy notices have become ever more prevalent on our phones and online. However, these notices are not well suited to their purpose of helping users make informed decisions. I suggest that instead of utilizing notice to eliciting informed consent, we could repurpose privacy notices to create the space for more meaningful, value-centered user decisions. Value-centered privacy decisions, or those that accurately reflect who we are and what we value, encapsulate the intuitive role of personal values in data privacy decisions. To explore how notices could be repurposed to support such decisions, I utilize Suzy Killmister's four-dimensional theory of autonomy (4DT) to operationalize value-centered privacy decisions. I then assess the degree that an existing technology, Personalized Privacy Assistants (PPAs), uses notices in a manner that allows for value-centered decision-making. Lastly, I explore the implications of the PPA assessment for designing a new assistant, called a value-centered privacy assistant (VcPA). A VcPA could ideally utilized notice to assists users in value-centered app selection and in other data privacy decisions.

**KEYWORDS**: DATA PRIVACY, PRIVACY NOTICE, PRIVACY ASSISTANT, AUTONOMY, VALUES

## 1 Introduction

Over the last four years, those of us browsing the Web from the European Union have become accustomed to a familiar sight – privacy notices, or "privacy pop-ups," asking us to agree to a website's privacy policy. These notices are a result of General Data Protection Regulation (GDPR), passed in 2016 and implemented in 2018 (General Data Protection Regulation (GDPR), 2016). Under the GDPR, informed consent is one of six legal bases[1] by which personal data can be collected and processed. This legal basis resulted in a flurry of privacy notices to elicit consent (Degeling et al., 2019).

While disclosures of this kind existed before the GDPR, the sheer volume of them has brought discussions surrounding the value and effectiveness of privacy notices back to the forefront. Privacy notices themselves are grounded in a long history of consent in other fields, especially bioethics.[2] Like consent forms for medical procedures or clinical trials (Beauchamp, 2011), privacy notices ideally aim to elicit informed consent from an individual.

Asking for consent has normative roots in respecting autonomy. Foundational documents of bioethics, such as the Belmont Report and the Declaration of Helsinki (K. J. Ryan et al., 1979; World Medical Association, 2013), stress the importance of respecting autonomy as a means to checking exploitative and manipulative practices.[3] James F. Childress, co-author of the Belmont Report, further purposes that respect for autonomy requires that we not only refrain from manipulating (a negative duty), but also that we disclose information and foster autonomous decisions (positive duties) (Childress, 1990). These duties have been operationalized into informed consent – that is, facilitating voluntary choice by providing information that is comprehensive and complete (Beauchamp, 2011).

---

[1] The six legal bases for processing data are: consent; contract; legal obligation; vital interests; public task; or legitimate interests. These are specified in Article 6 of the GDPR.

[2] There are special considerations to medical research that may not be applicable to data collection situations – eg, issues of dependency and vulnerability in healthcare situations introduce added consent concerns. There are, however, overlaps between bioethics and data ethics - especially as Big Data, smartphone apps, and artificial intelligence enter the medical field. For examples, see: (Jongsma et al., 2018; Klugman et al., 2018; Lucivero & Jongsma, 2018).

[3] The traditional catalysts for the Belmont Report and the Declaration of Helsinki were the Tuskegee syphilis trials, where black men were left with untreated syphilis and observed, and revelations surrounding Nazi experimentation on prisoners during WWII. See: (Brandt, 1978; Carlson et al., 2004).

For privacy notices to be effective mechanisms of informed consent, then, they must allow for voluntary choice based on comprehensive information. Discussions surrounding the "privacy paradox" phenomena – or the observation that the choices we make when faced with a privacy notice do not match what we say our privacy preferences are – suggests that neither of these criteria for informed consent are met (Spiekermann et al., 2001). There are two general arguments that have been used to explain the privacy paradox, one in line with privacy decision making as a rational, preference-based endeavor and the other postulating that humans are "flawed" thinkers who make questionable privacy decisions (Solove, 2020). The first argues that our privacy behavior, and not our stated privacy preferences, *truly* reflects how much we value privacy against, say, the good or service being offered. This is because we have been adequately informed by the privacy notice or policy and are acting in a rationale manner. In contrast, the second argues that our data privacy choices when faced with a privacy notice are not an accurate reflection of privacy's value but rather the result of human bias and heuristics – cognitive "tricks" or shortcuts that help us make fast decisions (Thaler & Sunstein, 2008; Tversky & Kahneman, 1974). Cognitive and behavioral psychology and its offshoot, behavioral economics, overwhelmingly supports the latter view – and, indeed, few would agree that we are purely rational *homo economicus*[4] when it comes to the decisions we make when faced with a privacy notice.[5] This suggests that not only are we not informed by privacy notices, but that we are susceptible to manipulation through exploitation of heuristic-based thinking. To explore these challenges in more depth, four examples of this "flawed" human thinking, followed by a brief look at the lack of comprehension problem, are explored below.[6]

To begin, there are *framing effects*. We are more likely to consent to data collection if it is framed in nonthreatening ways (O'Neill, 2002; Thaler & Sunstein, 2008). In addition, we can be encouraged to consent by utilizing the *inertia bias* – or, the human tendency to stay with the default condition (Thaler & Sunstein, 2008). For example, pre-selecting "Agree to All" in privacy notices increases consent to online data tracking (Utz et al., 2019).

We can also be coaxed into a false sense of security by utilizing the *representative heuristic* and/or the *conjunction fallacy* (Lewis, 2017; Thaler & Sunstein, 2008; Tversky & Kahneman, 1974). Privacy notices usually link to privacy polices – long, complicated documents packed with legal terminology (Jensen & Potts, 2004). We may believe something to be true because it matches our mental images of what it should look like – in this case, a long legal document resembles a "strict" privacy policy. This is the *representative heuristic*. Closely related to the representative heuristic is the *conjunction fallacy*, or the tendency to have our predictions misled when flooded with truthful facts (Lewis, 2017; Tversky & Kahneman, 1974). Privacy policies can be written to selectively reveal or flood us with truthful facts in a way that can mislead our predictions because they remain tactfully vague on other aspects.

Indeed, services providers that collect data can utilize these design strategies to coax us into making the choices they want us to make. These designs, called "dark patterns" or "deceptive design patterns", are ubiquitously used on privacy notices and have noticeable, measurable effects on data privacy decision-making (Brignull, n.d.; Gray et al., 2018; Mathur et al., 2019; Utz et al., 2019). To describe the consent given on these privacy notices as voluntary would be a stretch.

Now, on to comprehension. While it is difficult to measure and results vary, it has been suggested that as low as 0.24% of us read online privacy polices (Jensen & Potts, 2004). Practically speaking, no reasonable person *could* read all of them – one study estimated that it would take 244 hours to read all the privacy policies we see in one year alone (McDonald & Cranor, 2008). Like this challenge of *privacy policy fatigue*, too many privacy notices can cause *privacy notice fatigue*. This results in many us deciding to simply "click though" notices rather than reading them (Ben-Shahar & Schneider, 2010; Schaub et al., 2015). Disturbingly, a study of online social networking services estimated that this "click through" rate could be as high as 74% (Obar & Oeldorf-Hirsch, 2018). Eventually, this notice fatigue can cause us to become "apathetic users"- those who decide to consent every time to a service's data collection practices because they "no longer care" about their data privacy.[7] We are, in sum, overwhelmed instead of informed by privacy notices.

---

[4] Term taken from: (Hoofnagle & Urban, 2014). This text explores the influence and pitfalls of American privacy scholar Alan Westin's segmentation model, where a rational consumer reads privacy policies and makes decisions based on her preferences. There are three types of consumers according to Westin: "privacy pragmatists", "privacy fundamentalists", and "privacy unconcerned", with the pragmatist in the majority. According to Hoognagle and Urban, however, these "pragmatists" are hardly pragmatic at all, often making decisions with little or no understanding of the privacy protections in place.
[5] There is also discussion on whether privacy is something that can or *should* be tradable for a good or service. For example, see: (Allen, 2013).
[6] For a more comprehensive overview, see: (Solove, 2020).

These revelations from behavioral and cognitive psychology pose significant threats to the value of privacy notices. To address these challenges, research has been conducted to redesign and improve privacy notices to be more effective. Nearly every aspect of notice design – from timing, channel, modality, and control – has been extensively explored (Schaub et al., 2015). Particular interest has been paid to utilizing our heuristics to our benefit – that is, to "nudge" us beneficially. Nudging involves altering someone's environment in such a way that encourages one decision, without forbidding alternative decisions (Thaler & Sunstein, 2008). In the case of privacy notices, these nudges can be, for example, in the direction of better privacy (preserving) choices, or to increase comprehension/retention of a privacy policy (Almuhimedi et al., 2015; Calo, 2012; Kelley et al., 2013; Waldman, 2018). These interventions have ranged from presenting privacy permissions in a clearer format ("Privacy Facts") when downloading a new smartphone application (Kelley et al., 2013) to informing users how much their location has been used (Almuhimedi et al., 2015). These privacy-preserving nudges on privacy notices are sometimes called, to contrast them to dark patterns, "bright patterns" (Grassl et al., 2021).

By utilizing notices to encourage us to "preserve our privacy," however, bright patterns can undermine the original goal of privacy notices – to respect autonomy and prevent manipulation. This is due to the nature of nudges themselves. There remains a fundamental, ongoing discussion concerning whether nudges are inherently paternalistic or manipulative, linked to larger discussions around autonomy. Of particular interest is the concept of volitional autonomy – or that our actions should reflect our authentic desires. Volitional autonomy can be traced to analytical philosophers Harry Frankfurt and Gerald Dworkin. According to Frankfurt's original theory of free action, to act freely – autonomously – we must act in accordance with our second-order volition, or, in other words, we must desire what we desire (Frankfurt, 1971). Dworkin similarly argued that autonomy is our ability to reflect and endorse "first-order" desires in accordance with our "higher-order" values (Dworkin, 1988b). Because those using nudges encourage the actions they desire the person nudged to desire, and not what the person being nudged necessarily desires to desire, the person nudged may act in a manner that they do not endorse (Hausman & Welch, 2010; Schmidt & Engelen, 2020). Put a bit more clearly, bright patters encourage us to preserve our privacy by making us want to preserve it, even if we don't "truly" want to. This violates our autonomy.

This concern around nudges and user autonomy more generally has been raised in other fields. In the field of psychology, for example, a similar conception of autonomy to that of Frankfurt's and Dworkin's has been operationalized by psychologists Richard Ryan and Edward Deci into their Self-Determination Theory (SDT). SDT defines autonomy as self-endorsement of an action according to one's values (R. M. Ryan & Deci, 2004). Empirically, studies utilizing SDT's Internal Locus of Causality, or I-PLOC, measurement suggest that even small nudges can undermine someone's autonomy (Arvanitis et al., 2020). This has implications for bright patterns, whose use – however well-intentioned – could cause an I-PLOC shift, again violating, rather than respecting, autonomy.

We are stuck at an impasse when it comes to how to design and utilize privacy notices. Insights from behavioral psychology suggest that informed consent via notice is not plausible. Instead, we are at worst, uninformed, and at best, overwhelmed when encountering a notice. We are manipulated and framed to "consent" our data away. Some have aimed to use privacy notices to nudge us to make "better" privacy choices, but this has possible negative implications on our individual autonomy. Perhaps, as many scholars argue, we should not use privacy notices at all.

---

[7] There is a caveat here that is worth mentioning: The term "apathetic user" is meant to capture those who would prefer to be more data protective but feel overwhelmed by notices to the point of "no longer caring," or apathy. Someone who reflects upon their data privacy choices and decides to give their data away "always" because they, say, believe that sharing their data will stimulate technological progress and improve their user experience are not meant to be captured as apathetic users. However, I would argue that few of us "click through" notices because we believe in what we are doing, and more so because we feel overwhelmed.

I don't believe this is the right question to be asking. Instead, we could try to imagine a role for privacy notices outside that of eliciting informed consent. Perhaps it is not the notices themselves that are the problem, but a mismatch between the task and the tool. This argument has been made by Daniel Susser (2019), who suggests that privacy notices could perhaps promote individual decision-making by increasing awareness of the current data privacy landscape; helping identify potential privacy concerns; and assessing our legal rights, where applicable.[8]

In this paper, I would like to take this one step further, by marrying Susser's intriguing idea to explore notice outside traditional notice-and-consent to notice's original goal of respecting autonomy. While we must not prioritize respect for autonomy above all else and still balance it with other values relevant to data privacy - values such as accountability, transparency, and trust (O'Neill, 2020; Waldman, 2015, 2018)[9] - respecting autonomy is still a worthwhile aim for data privacy decisions. Like in bioethics, respecting autonomy in data privacy decisions still provides a critical check to exploitation and manipulation despite informational power asymmetries – it is, at the very least, the first line of defense (Susser, 2019; Susser et al., 2019). There is also a sense of disempowerment and an increasing learned norm of simply "giving up" on protecting our data privacy. In the case of those who reside in liberal democracies, this disempowerment is fundamentally out of synch with the central tenants of our governments.[10]

To respect autonomy, then, we could try to utilize the insights of behavioral psychology and accounts of autonomy in philosophical literature to operationalize autonomy in a manner that doesn't rely upon informed consent. To be clear, I am not aiming to make regulatory recommendations or assess the GDPR, which, as the previous paragraphs demonstrate, likely will need to atone to the lessons of behavioral psychology just like notices. I leave this task to other scholars.[11] Instead, I am intrigued by Susser's call to find other uses of notices beyond traditional informed consent.

I am particularly interested in conceptualizing and operationalizing autonomy in a manner that better captures the intuitive role of personal values in data privacy decisions - what I call *value-centered* privacy decisions. This is because re-purposing notices for value-centered privacy decisions could retain the original normative underpinnings of notice – to respect autonomy – while also promoting more meaningful data privacy decisions and an overall better experience. Because of this emphasis on personal values, I will also not be exploring here the collective implications of privacy disclosure or the social value of privacy.[12] Similarly, value-centered privacy decisions is not meant to encapsulate discussions around broader, public values that may be relevant for governments when constructing policies around data disclosure.[13] . I do not intend to dismiss these discussions, but rather to complement them with one re-focusing on the individual experience of data privacy choices – those small, often unsatisfying privacy decisions that we make daily as we interact with digital technologies. In addition, once we operationalize autonomy for notice using value-centered privacy decisions as a starting point, we could utilize dynamic privacy technologies to translate this into practice. We could, in other words, instead of designing for more *privacy-protective* choices, design for *value-centered* privacy decisions s.

---

[8] See: (Susser, 2019) pg. 38: "If the problem with notice-and-consent as a whole is that it fails to facilitate and respect individual agency over data, then we ought not to deprive ourselves of even flawed and partial mechanisms for strengthening such agency."

[9] Trust – and placing it well - is a pressing concern for data privacy decisions. For example, in (O'Neill, 2020), Onoara O'Neill notes a number of problems with trust and accountability in the digital age. Technology allows new intermediaries to control online content opaquely, and it is difficult to decide whether or not they are trustworthy. This results in unrealistic accountability mechanisms, like data privacy policies and excessive notices.

[10] I particularly appreciate how this is stated in (Susser et al., 2019), pg. 8: "Autonomy is in many ways the guiding normative principle of liberal democratic societies. It is because we think individuals can and should govern themselves that we value our capacity to collectively and democratically self-govern."

[11] This has been done quite well elsewhere – in particular, Frederik Zuiderveen Borgesius devoted an entire PhD to this subject. While this was completed in pre-GDPR (EU Data Protection Directive) days, many of his analyses regarding informed consent are still relevant to the GDPR. See: (Zuiderveen Borgesius, 2015).

[12] For an example of a discussion exploring the social value of privacy, see: (Roessler & Mokrosinska, 2013).

[13] Examples of these discussions can be found around COVID-19 tracking applications. The purposed benefit to public health was considered in terms of efficacy and uptake of the apps (Luciano, 2020; Morley et al., 2020); justice, equity, and solidarity with vulnerable subpopulations (Hendl et al., 2020); and civil liberties and surveillance (Kitchin, 2020), to name a few.

To this end, I will firstly define what is exactly is meant by "value-centered privacy decisions." To accomplish this, I will draw from interdisciplinary literature as well as invoke our own intuition. I will then link value-centered privacy decisions to autonomy– in particular, Suzy Killmister's Four-Dimensional Theory of Autonomy (4DT) (Killmister, 2017). After providing this link, I operationalize 4DT within the context of individual data privacy decisions.

In the second half of the paper, I move on to how we can use notices to create a space for value-centered data privacy decisions. To do this, I will utilize 4DT to help design a system – called a Value-Centered Privacy Assistant (VcPA) – that could help create that space for value-centered data privacy decisions using privacy notices. The VcPA would constitute an extension of an existing technology, called personalized privacy assistants (PPAS), which are machine-learning systems that provide personalized privacy notices (Liu et al., 2014, 2016; Warberg et al., 2019). In particular, I will explore how to design a VcPA to help users when they are deciding whether or not to download an app on the Apple or Google Play Store. To do this, I will utilize 4DT to assess to what extent value-centered decision-making is supported by existing PPA technology and identify areas of improvement to consider when designing a VcPA. I conclude that a VcPA with these modifications could effectively utilized notice to assists users in value-centered app selection and in other data privacy decisions.

## 2 Notice revisited: promoting autonomous value-centered privacy decisions

Privacy notices do not help users make informed privacy decisions. In this section, I explore how they could instead be utilized to create the space for us to make autonomous, value-centered privacy decisions – designing for autonomy, rather than for privacy, in privacy decision-making.[14] I firstly outline how values are involved in privacy decisions before detailing its link to autonomy.

### 2.1 Individual data privacy decisions (ideally) reflect user values

Value-centered privacy decisions are a means of capturing the intuitive role of personal values in data privacy choices, with the aim of making the individual experience more meaningful. Value-centered privacy decisions can be understood as those that result in ends that accurately reflect our personal values. Consider your home – the furniture you have there; the art you put on the wall; and who you let in should ideally be a reflection of who you are and what you value. We can imagine, however, instances when this won't be the case – perhaps your local furniture store doesn't carry the color of furniture you like and other stores are too far away for you to reasonably travel to. Similarly, your smartphone is like your "digital home" – what apps you download and who you allow to access your data should ideally reflect your values.[15] However, hurdles such as dark patterns may restrict your privacy choice to the point that it may not reflect your values. Individual privacy decisions can therefore be evaluated to the extent to which they result in ends that accurately reflect the individual's personal values.

---

[14] Exploring ways of promoting autonomy in technological design is not a new concept. It has, for example, been explored in the field of human-computer interaction (HCI) and is a central component of value-sensitive design (VSD) (see: (Peters et al., 2018) for HCI and (B Friedman & Nissenbaum, 1997; Batya Friedman, 1996). My approach, however, is slightly different - it links personal values to data privacy decisions, which in turn, is linked to autonomy. This is explored more in section 2.

[15] Privacy "the value" is here understood as an instrumental value, valuable for other terminal values that making a privacy decision (ie, being private or not) brings about. This allows us to consider not only traditional values associated with privacy-preserving behavior (such as security), but also other values that may be associated with sharing data (such as sharing photos on Facebook to keep family and friends "up to date" on your life).

Many disciplines have explored this relationship between privacy and values more broadly, albeit in different contexts and with different emphases, and I have drawn from this multidisciplinary literature to inform this conceptual link of individual data privacy decisions and personal values. On the legal side, Daniel Solove has conceptualized privacy (more generally) as many related items that can be encompassed under a common heading without necessarily having a single theoretical basis (Solove, 2002, 2007). He further puts forth that privacy is instrumental in that it allows an agent to protect or promote valuable ends (Solove, 2002). Seeing individual privacy decisions as value reflection retains his idea of both value plurality and instrumentality. Here, however, value reflection emphasizes less the ends that are brought about than which personal values are ultimately reflected by the user achieving those ends.

Values as ends in themselves is drawn partly from Value-Sensitive Design (VSD), which proposes that technology embeds and expresses values (Batya Friedman et al., 2008). I say in part because an individual's data privacy decision has consequences that could be seen, ideally, as a reflection of her *personally-held* values, rather than greater stakeholder values that concern a designer utilizing a VSD approach; it is, one could say, VSD with an emphasis on the individual experience. Value-centered privacy decision-making also incorporates an aspect of *information flow,* or, in this case, the data that is being collected as the result of user's privacy decision. This a central idea of philosopher Helen Nisssenbaum's contextual theory of privacy (Nissenbaum, 2004), which puts forth that privacy can be understood as *appropriate information flow* that match the context-specific norms. Here, again, value reflection is not interested in generally consensual norms but rather the individual's specific values and the degree to which they are accurately reflected following a data privacy decision.

2.2 Ideal value reflection (and designing for it) requires autonomy

As stated in the introduction, designing for value-centered choices links back to the original goal of notices – respect for autonomy by designing for it. This is because value reflection cannot occur if autonomy is frustrated. In order for value-reflection to be "ideal" – that is, an accurate reflection of someone's values – a privacy decision must be sufficiently unhindered by external forces such as dark patterns. She must also be sufficiently engaged with the privacy decision to make a conscious, value centered decision. We are, essentially, designing for autonomy – with the goal of providing more meaningful privacy decisions that reflect who someone is and what she values.

Furthermore, if we wish to explore ways of utilizing notices in a manner that supports value-centered data privacy decisions, we require a means of assessing to what extent one's data privacy choices meets this "ideal". We need, in sum, to define and identify conditions for autonomy in data privacy decisions.

To do this, we can look to the wealth of autonomy literature in philosophy to identify a concept autonomy that fits our context – individual data privacy decisions. To fit our purposes, the conception must meet three criteria.

Firstly, it must have *personal values* at its core to be harmonizable with the idea of value reflection.

Secondly, it must be *reasonably systematic* and practical in order to assess data privacy decisions.

Thirdly, because we are looking at individual data privacy decisions, it must be a theory of *personal autonomy* that is still able to account for the apathetic user phenomena (described in the introduction). We are not aiming to assess how the decision whether or not to disclose our data (our data privacy decision) affects our overall autonomy or greater democratic processes, but rather, how autonomous the decision to disclose our data is.[16] In addition, we are not investigating whether what someone values and acts on in her privacy decision is the most morally justifiable one, but whether it is a reflection of her values. We are, therefore, not looking for a conception of moral or Kantian autonomy, but one of personal autonomy.[17] At the same time, there must be some normative incentive to *not* become an apathetic user by succumbing to notice fatigue – otherwise, there will be no opportunity to make a value-centered choice. This final requirement – with all its nuances – is the trickiest of the three to meet.

---

[16] For examples of these discussion see: (Cohen, 2013), who discusses the role of privacy in setting boundaries from external influences in order to define who we are and what we want; or (Zuboff, 2019), who explores the danger data disclosure and behavioral modeling pose to our agency and the greater democratic process.

Suzy Killmister's Four-Dimensional Theory of Self-Governance (abbreviated hereafter as "4DT") meets these criteria. 4DT divides autonomy into four dimensions – *self-definition*, *self-realization*, *self-unification*, and *self-constitution* (discussed in greater detail later in this section) (Killmister, 2017). To the first criteria, self-definition has links to values – it involves forming commitments on how to be and act in the world which, in turn, are clustered into values. 4DT also meets the second criteria: it categorizes autonomy into four distinct, accessible, and practical dimensions.

4DT can also account for the final, and most difficult, requirement – it must be a theory of personal autonomy that also discourages apathetic choice. To the first criteria, 4DT is indeed a theory of personal autonomy – it is not concerned with whether the ends that are brought about are necessarily the morally correct ones. To the second criteria, it can also capture the "apathetic user" phenomenon. Normativity for 4DT is derived from an agents' ability to take on or reflect upon their own commitments – that is, we must be committed to something we should do or become (Killmister, 2017). These agency requirements are captured in *self-constitution*. Thus, someone a who refuses to take on any commitments or intentions whatsoever – such as an apathetic user – cannot be said to be highly self-constituting. Fitting this last part can be a challenge for closely related theories of personal autonomy and is, ultimately, where 4DT stands out. Returning to Dworkin, for example, if the "apathetic user" decides to take on no commitments concerning privacy decisions after the highest order of reflection, it could be considered autonomous (Dworkin, 1988a).[18]

Having established that 4DT meets the criteria for exploring value-centered data privacy decisions, we can now examine it in more depth; operationalize its dimensions in the context of data privacy decisions (summarized in Table 1); and ultimately, in the next section, utilize it to re-purpose privacy notices to promote value-centered privacy decisions.

As mentioned above, in her book, *Taking the Measure of Autonomy: A Four-Dimensional Theory of Self-Governance* (2017), Killmister maps the complex landscape of autonomy into four different dimensions. The first dimension, *self-definition,* is concerned with personal identity: self-definition assesses the level of internal consistency between the goals, beliefs and values that make up our personal identity. While self-definition is not particularly relevant to the *act* of privacy decision-making, it does help us better visualize how values are involved and reflected in the privacy decision-making process. As outlined in Figure 1, our self-defining attitudes – beliefs, goals, and values – commit us to act a certain way.[19] For example, perhaps I believe too much screen time is bad for my health. I am therefore committed to spend less time on the computer. Similar commitments can then be clustered together into values. Perhaps I also have the goal of running a marathon. This commits me to run every day, which could be clustered with less computer time to say I value my health.

The second dimension, *self-realization,* is concerned with practical agency: self-realization assesses the level to which our conclusion from practically deliberating aligns with our intentions (*internal self-realization*), and the degree to which our actions align with our intentions (*external self-realization*). Returning to our example, let's say I am considering downloading a running smartphone app that would like to access my health data. I deliberate and determine based on my values and, because I value my health, conclude that I best ought to allow it access. I form the intention, or in this case, a *privacy preference*, to allow the app access to my health data. I then act on this and allow it access – that is, make a *privacy decision*. This upholds self-realization, as I deliberated, formed an intention, and acted coherently.

---

[17] This is not to say that privacy decisions cannot have moral weight, but merely that it is not what we are investigating here. For an example of moral duties towards privacy disclosure (Allen, 2013), where she draws on Kant to argue that we have a duty toward ourselves to protect our privacy out of respect for ourselves.

[18] There are, however, instances where someone may act in a manner *similar* an apathetic user while still being self-constitutive. For an example, please see footnote 7.

[19] Self-governing policies can also influence self-defining attitudes and have downstream effects on out privacy preferences. Self-governing polices are that dictate what and how to "believe, plan, and value" (Killlmister, pg. 22).

The final two dimensions are concerned with the relationship between personal identity and practical agency. *Self-unification,* whether our actions are consistent with their personal identity; and *self-constitution,* our ability to take on and form commitments. In the case of downloading the health app, self-unification is upheld because my action to download the running app and give it access to my health data is consistent with valuing my health. Self-constitution is also upheld because I am able and willing to form conclusions and intentions concerning whether or not to download the app.

Applying 4DT to this context - individual data privacy decisions - comes with a few notable caveats. These concern self-unification and self-constitution. Firstly, self-constitution originally pertained to someone's *overall* willingness to take on commitments. She could, then, still be considered highly self-constituting if her willingness is present in a *range* of areas, even if this willingness does not extend to *all* circumstances or topics. In this case, I am utilizing 4D as a tool to analyze autonomy constrained to a specific circumstance – individual data privacy decisions – and am treating self-constitution as *the ability to take on any or new commitments pertaining to data privacy decisions*. Secondly, self-constitution can also apply to both the domains of personal identity and practical agency. It encompasses the degree to which someone forms self-defining attitudes; the degree to which she can practically deliberate and form intentions; and the degree to which she is able to coordinate the two. When considering how self-constituting she is with respect to data privacy, self-defining attitude formation and its (continued) coordination with practical deliberation is a longer-term process and not one that is made at any one privacy decision point; thus, self-constitution in terms for our purposes will be primarily concerned with the degree to which someone is able to practically deliberate and form intentions. This brings me to the third and final caveat: self-unification is not only concerned with ensuring that one's actions match their personal identity, but also the extent to which the conclusions/commitments generated during practical deliberation can inform transformation (a change in personal identity). Again, this is tied into self-defining attitude formation and is not something that is made at one point in time. Because of this, an analysis of data privacy-decision making in terms of the self-unification dimension will focus on the overall coherence of someone's actions to her personal identity.

# 3 Personalized privacy assistants technology for value-centered privacy decisions

So far, we have identified what value-centered privacy choices are; linked them to autonomy; and now, we can start designing for value-centered autonomous decisions using privacy notices. In this section, we will utilize the dimensions of 4DT to systematically assess the degree to which current Personalized Privacy Assistants (PPAs) use notices in a manner that creates space for privacy decisions that reflect our values (summarized in Table 1). In addition, we will also be mindful of insights from behavioral psychology that have undermined the value of notice in the past, harmonizing them with 4DT where applicable. This analysis of PPAs is done to inform the design of a value-centered privacy assistant (VcPA) for smartphone app selection). Privacy assistants use dynamic and selective notices to improve the user privacy decisions, providing us with an excellent starting point for exploring notice use to create value-centered privacy decisions and eventually conceptualize a VcPA.

## 3.1 A note on context

I will firstly assess smartphone PPAs as they currently exist – that is, assisting users with privacy settings *after* downloading the application. I will then utilize this information to inform the design of a VcPA to assist a user with one specific privacy choice – when selecting a smartphone application. While there are many different individual privacy decisions that could be explored and could benefit from value-centered autonomy-enhancing notices, it it is not feasible to explore them all.The focus of this paper is when users are deciding to download smartphone applications on the App or Google Play store, and there are a few reasons why this particular privacy decision point is of interest.

While it has been argued that a per-app permission basis is not a helpful context to explore for privacy interventions,[20] there are several reasons why looking at the decision whether to download an application is appealing and could complement existing "post-download" interventions. To start, there is something to be said about empowering us to manage data collection in line with our values by preventing our data from being collected in the first place. While agreeing that the centralization of data collection into a select few third-party libraries cannot be combated with the current permission choices available to us, targeting the decision to download could take into account this third-party sharing issue by preventing value inconsistent data sharing to a library in the first place. It also provides an extra layer of control over our personal data that could further complement – rather than contradict – other initiatives. Improving control in a manner that increases value reflection at the decision whether or not to download an application could also increase market pressure for improved products that better align with our values.[21] In this manner, value-centered application selection could serve as the first line of defense against value-inconsistent data practice and encourage more consistent applications to be made in the future.

## 3.2 Evaluating Personalized Privacy Assistants (PPAs)

We can now turn to Personalized Privacy Assistants (PPAs): what they are, and why explore them as a means of facilitating value-centered smartphone app selection choices. Briefly, PPAs, currently under development by a team at Carnegie Mellon University,[22] would be machine learning assistants that personalize and automate privacy choices for a user. The team has explored PPAs to help manage user privacy in the Internet of Things (Das et al., 2018) and smartphone applications (Liu et al., 2016), to name two. Personalized assistants could also utilize a variety of approaches to help a user manage their privacy, such as privacy-preserving nudges or semi-personalized setting recommendations based on user preference profiles (Liu et al., 2014; Story et al., 2020; Warberg et al., 2019). For example, Liu and colleagues (2016) designed a smartphone PPA where users got personalized recommendations for their privacy controls on their Android smartphone. They firstly developed privacy preference profiles based on a dataset of Android user privacy settings. During a user study of the PPA, participants took a dynamic, short quiz on their privacy preferences to sort them into a privacy preference profile. Based on their profile, they were then given recommendations on how they could change their privacy controls on their phone.

Conceptually, the development of PPAs has been fueled by a desire to help users make the best privacy choices for them in the current digital privacy environment. In particular, PPAs seem to be focused on overcoming one aspect of this larger issue – user notice fatigue – something an intervention for smartphone application selection would also want to avoid (Liu et al., 2014). As Florian Schaub and colleagues have previously described (Schaub et al., 2015), determining the proper amount of notices is exceedingly difficult – too many notices causes users to simply "click through" them, and deploying them at too little a frequency doesn't provide the user with adequate information to make an informed decision. Through dynamic, personalized recommendation, PPAs could overcome the notice fatigue concern by only presenting notices to the user that are relevant to them. This makes PPAs an attractive technology to help us make more autonomous, value centered privacy decisions, where notice fatigue is huge barrier. By utilizing 4DT, we can explore PPAs to assess to what extent they are successful at dealing with the issue of notice fatigue by addressing the 4DT failures that constitute notice fatigue. We can also use 4DT to identify areas of improvement to inform the design of a value-centered privacy assistant (VcPA), a system to help use make app choices in a manner that best reflects our personal values.

---

[20] For example, Chitkata and colleagues have argued that looking to improve per-app permissions is unproductive because 70% of app data access requests are by the same 100 third party libraries (Chitkara et al., 2017). Instead, they created ProtectMyPrivacy – an application that provides users with the ability to control which libraries as well as apps have access to their data. In addition, cross-app tracking has also been seen as a major an issue, leading to Apple to release feature of "App Tracking Transparency" feature, which requires smartphone apps to get permission from users to track their activity across many applications, with iOS 14.5 (Apple, 2021).
[21] In (Susser, 2019), Daniel Susser makes a similar argument – privacy notices that disclose a company's data practices – even if incomplete – could encourage them to meet social norms concerning privacy.
[22] Also see: The Personalized Privacy Assistant Project (https://privacyassistant.org).

3.2.1 PPAs and Notice Fatigue

Not surprisingly, from the standpoint of 4DT, PPAs are mostly successful at addressing notice fatigue by utilizing *selective* (rather than blanket) notifications (Liu et al., 2014, 2016; Warberg et al., 2019). The phenomena of notice fatigue can be understood as a failure to self-realize, self-unify, self-constitute, or a combination of these. We can consider failures to self-realization, where a someone "just clicks through" *all* notifications despite an intention not to; self-unification, where this "click through" action does not match her personal identity and her values; and, in more extreme cases, self-constitution, where an "apathetic user" becomes so overwhelmed by privacy notices that she stops forming new intentions and commitments about her data privacy *at all*. In addition, challenges arise when two values come into conflict with each other. For example, many of us, for a variety of reasons, value efficiency. We can, for example, imagine a someone who values self-determination, control, and efficiency. In the current privacy notice landscape, her values will be a odds with each other – she will be in a *double bind*,[23] and no matter how she acts, she will act contrary to her values (fail to self-unify).

      PPAs also act as a form of self-binding, that is, holding the us to follow through on her intentions. They do this by utilizing selective notifications to prompt her to re-consider her privacy decisions that are at odds with her preferences. This, in turn, may help her act more consistently with her personal identity. By selectively "slowing down" the user in a manner tailored to her individual privacy preferences (Kahneman, 2011), she has further opportunity to pause and re-consider an intention that is at odds with the conclusion of her practical deliberation.[24] This "slowing down" can be considered a beneficial form of friction, or disruption of a goal-focused mindset. While friction was originally considered something to be minimized in technological design, its positive aspects such as its link to slower, more mindful user thinking have received more recent attention (Cox et al., 2016).[25] In the field of data privacy, friction has been proposed as a means of encouraging reflection and attention to one's own underlying beliefs and values when engaging in a data privacy decision (Terpstra et al., 2019). Friction in the case of PPAs can therefore be understood as a form of self-binding that helps the user self-realize and self-unify by triggering mindful reflection.

      PPAs also introduce this friction in a manner that is more observant of what the user desires. I would argue that over-generous or designer-selective use of friction can cause problems. While these cases would encourage user reflection and, by extension, autonomous choice, the challenge here is that the *person designing the friction decides* when the user should slow down and be reflective, rather than when it would be most beneficial to the user to do so. This could either result in the user only making conscious decisions when the designer thinks that she should, or a well-intention designer may use friction too liberally as to cause notice fatigue. By selectively notifying users *based on what she thinks is best for her*, PPAs are theoretically better at assisting users without slipping into these issues.[26]

      Considering now the challenge of double binds, the PPA does help alleviate tensions between values – but only to a certain extent. In terms of a double bind concerning efficiency and other values

---

[23] I am using Killmister's terminology here, from Part II: Applications of (Killmister, 2017), where she explores, among other aspects, the autonomy of the oppressed, and how they are often forced into double bind situations.

[24] Here, by "slowing down" I mean the shift from fast to slow thinking as described by (Kahneman, 2011). "Fast" System 1 thinking is automatic, mindless, and ripe with heuristics and biases. "Slow" System 2 thinking is conscious, deliberate, and mindful.

[25] Empirical studies have supported this idea, such as: (Mejtoft et al., 2019), where well-placed friction to increase mindfulness when interacting with a mobile application also increased user satisfaction.

[26] However, as a machine learning assistant, there are a number of concerns that PPAs need to consider such as minimizing bias and being sufficiently transparent. I discuss these issues in Section 3.4 when considering designing a PPA-like system for app selection, called a value-centered privacy assistant (or VcPA).

described previously, she may receive an excessive number of selective notifications and there may be instances where she receives a notification that is not needed, and therefore, inefficient, given the large number of different possible privacy preferences and situations. Ideally, the notice's selectivity would be tuned to not overwhelm the user and lead to apathy, but the number of privacy preferences and situations could make this tuning exceedingly difficult. Thus, while PPAs can help with double binds in part – a better solution may be possible.

3.3.2 – Other challenges: inertia bias and privacy controls

Similar to the issue of notice fatigue is the inertia bias. As mentioned in the introduction, this cognitive bias makes it difficult for users to change from their initial privacy settings – even if they wish to (Thaler & Sunstein, 2008). In the case of PPAs, this manifests as a failure to update and change their privacy profile, even if they are given the ability to do so. Returning to the Liu and colleagues PPA paper (2016), researchers used nudges to test if users would change their profile. Most did not. In addition, while users were always able to change their profile, few did. While the authors claimed that this supports the accuracy of their profiles, this could also be interpreted as evidence of the inertia bias. We can also imagine someone who continued to receive notifications based on her original privacy profile and acted according to them even if it she determines that this profile is not the best fit for her. Like certain manifestations of notice fatigue, failing to act on this intention would result in a failure to externally self-realize. In addition, her resulting privacy decisions would likely not be consistent with her values, a violation of self-unification. As long as PPAs do not account for the inertia bias, they cannot be said to be fully self-realizing or self-unifying.

     Lastly, there is the challenge at the level of privacy controls themselves. There are a plethora of different privacy notice designs with different degrees of privacy control (Utz et al., 2019). Smartphone PPAs can also only modify preferences using the smartphone operating system's (OS) available controls. These controls (eg, access to Contacts, Camera, Location) have been previously shown to be insufficient for capturing the privacy concerns of users (Felt et al., 2012). In addition, Terpstra and colleagues (2019) have purposed that a lack of meaningful privacy controls in conjunction with positive friction can lead make us frustrated, possibly undermining positive friction benefits such as value reflection. This frustration can be understood in terms of failures to self-realization and self-unification. If a user forms and intention (privacy preference) but the control is not present, this would be a frustration of self-realization. Because she is not acting according to her values, this would also constitute a failure of self-unification. Due to the limited level of granularity available to them, our ability to realize our commitments and make choices consistent with our commitments is still stifled with PPA use.

3.3 In the context of selecting smartphone applications

These challenges present three major implications for designing a PPA-like system, a value-centered privacy assistant (VcPA), to help users select smartphone applications. Firstly, like current PPAs, users of a VcPA may fail to update and change their privacy profile due to the inertia bias, hindering self-realization and self-unification. Secondly, these dimensions could be further undermined due to the lack of granularity problem. While the challenges to self-realization and self-unification of current PPAs are a result of the lack of granularity of either 1.) privacy notice design (online PPAs) or 2.) the OS' privacy setting controls, a PPA-like system assisting with smartphone application selection is essentially a dichotomous decision: either download, or not download. Not only does this minimize self-realization, but ithis will prevent a user from realizing her commitments in a manner consistent (unifying) with her values by introducing additional double-bind situations. For example, consider a user deliberating whether to download a social media app such as Instagram. While she greatly values her social connections, she also values control over her life. She must decide whether to join social media and allow her data to collected or to abstain, neither of which will be in full accordance with her values., Thirdly, the aforementioned double bind situation concerning efficacy and PPAs will also need to be considered when selecting an app; someone who values efficiency may still face instances of double binds if selective notifications are based on the app's data collection practices and her privacy preferences alone.

## 3.4 Suggested modifications when developing a value-centered smartphone privacy assistant

There are a few possible modifications to current PPA design that could overcome these remaining challenges and create a value-centered privacy assistant (VcPA) (Table 1).

To rectify the efficiency value challenge, VcPA profiles could instead be based on the user's personal values. By basing profiles on values rather than privacy preferences alone, the VcPA could be more accurately tuned to prevent efficiency-based double bind situations and maximize efficiency. . User tests, however, will be required to 1.) determine in what way values intersect with app data collection preferences and 2.) how this intersection could be operationalized as VcPA profiles. In addition, the profiles must also: 1.) be an *accurate* reflection of a user's values; 2.) have an accessible and *understandable description*; 3.) clearly state *how* the privacy preferences and values are utilized in profile creation and assignment; and 4.) be able to be *changed* to a different profile if notices are unhelpful. All of these considerations must be accounted for when designing and testing a VcPA protype in order to rectify the double bind challenge without sacrificing accuracy and transparency.[27]

Self-realization and self-unification as defined by 4DT are minimized in part because PPAs are limited by the lack of relevant controls, which remain relevant to VcPAsthrough a download-or-not-choice that may make acting consistent with their values not be possible. This challenge could be overcome in a VcPA by not only informing the user when an application is requesting privacy settings *inconsistent* with their values, but by also suggesting alternative apps to quickly link them out to similar applications that better align with their commitments. This increases the likelihood that they will be able to find an app whose data collection practices match their values, thus (better) upholding self-realization and self-unification.[28]

In order to be more fully self-realizing and self-unifying, however, a VcPA will also have to tackle the inertia bias. Agreeing with others that there is a need for "learning" (or, at least room for change) in the privacy notice process (Terpstra et al., 2019), this could be tackled by periodically (but not excessively) "mining" user goals, values, and preferences – that is, checking in with the user that the notices are still a good fit for them by making them aware, perhaps by using an "exploratory notice." It will be critical, however, that this "mining" process be sufficiently random and dispersed as to *not* encourage user action one way or another. If a user makes a choice that is inconsistent with what they have resolved to do as a result of this exploratory notice, this would be inconsistent with external self-realization (and possibly self-unification if the action goes against her values). In the recommender system literature, researchers have empirically studied how to limit the effects recommendations have in encouraging certain behavior, suggesting that a process of "continuous random exploration" could be a way to combat these phenomena (Jiang et al., 2019). A similar approach to continuous random exploration could be utilized to time exploratory notices. User tests will be required, however, to determine the right balance between providing users with the ability to learn and explore through exploratory notices and inadvertently nudging them in a non-autonomous direction.

---

[27] To this aim, a protype VcPA has been designed and has been tested with users. For more on the technical development of the VcPA and its testing environment, please see (Carter et al., 2022)). At the time of writing, the research team is analyzing participant feedback and will report the findings in a forthcoming paper.

[28] An important caveat here is the case of social media applications such as Instagram, mentioned in Section 3.3 as presenting a possible double bind situation. A VcPA that recommends alternative applications will only be effective at addressing the double bind and self-realization/unification concerns if such an alternative is available. This may not be the case for social media and messaging applications. Returning to the example user that values both social connectivity and control and is deliberating whether to join Instagram. If most of her friends and family are on Instagram, an alternative with similar social value *to her* that is likely not available. As a VcPA can only work within the existing app ecosystem, including social media apps that essentially have "monopolies" on online socialization, they are unlikely to be effective at tackling double binds in this context. This may require regulatory, "trust-busting" interventions, a broader discussion of which is out of the scope of this work. However, I would argue the potential for VcPAs to promote value-centered decisions within other app contexts makes them still worthy of exploration.

## 4 Final Thoughts

Here, I have proposed a new goal for privacy notices – to create the space for value-centered privacy decision-making. This understands data privacy decisions as, ideally, a reflection of user values. After linking this to autonomy, I then explored how we can design for this ideal situation by designing for autonomy – that is, *create the space* for value-centered privacy decisions – using 4DT and remaining mindful of insights from behavioral psychology. To inform the design of a smartphone assistant that creates this space for users, I first examined PPA technology using a 4DT lens. While PPAs, with their use of selective notices, are partly successful at creating the space for value-centered choice, a number of concerns around the inertia bias, double bind choices, and lack of controls remain., . Using these insights to inform notice timing, content, and basis, I lastly propose a value-centered, smartphone privacy assistant, (VcPA) to help users make more value-centered decisions at one privacy decision point: smartphone app choices.could

There are, however, varies concerns of this approach that still need to be addressed. One aspect of PPAs in general –whether value-centered or not – that has not fully been addressed here is that the collection of data itself for running such a system could be utilized to undermine a user's autonomy. For example, in a value-centered system, information concerning a user's personal values and how they relate to their app choices is quite sensitive in nature and could be utilized by targeted advertisers to nudge users towards products more effectively. I think this is, however, fundamentally a technical challenge than a foundational threat to the VcpA concept. It is an issue of data protection that could be addressed by ensuring that the data required to run the VcPA is held locally on the user's device and able to interact with an app store without disclosing any information about the user (Carter, 2021). Local data storage, however, could take up too much space on a user's phone and make a VcPA impractical. These challenges will have to be addressed during the technical development of the assistant, in addition to determining how to design VcpA profiles based on personal values and the ideal frequency for the exploratory notices.[29]

In addition, empirical studies will be required to validate the conceptualization of data privacy decisions as a reflection of personal values and to assess whether the VcPA notices designed to promote them are effective. To this end, a mixed-methods study is currently in progress. The main aims of this study is to determine if 1.) a VcPA system is beneficial over traditional smartphone privacy notifications and controls; and 2.) if such a system assists users to make app choices more consistent with their values. More fundamentally, this study also aims to assess the role of values in privacy-decision making and, in particular, smartphone application selection. Answering these questions will be critical to determining the feasibility and desirability of a smartphone privacy assistant that enables value-centered privacy decision-making.

Finally, VcPAs should not be the only mechanism by which we govern data flows and manage privacy. While I have suggested here that a VcPA could increase value-centered privacy decisions by re-purposing notices, data privacy decisions have broader implications than the individual who makes them. Individual data flows can, for example, be grouped into profiles by "big data analytics," creating a group of algorithmically-related individuals with interests that go beyond that of any one group member (Mittelstadt, 2017). While I would disagree that these collective properties of data privacy means that respect for autonomy – the normative basis for value-centered privacy notices - has become null and void, I recognize that an autonomy-exclusive solution does not accurately capture all the dimensions of data privacy. I also want to acknowledge genuine concerns about the effectiveness and desirability of our current privacy regulations. For example, initial empirical data suggests that smartphone data tracking on Android devices has not significantly changed following the introduction of the GDPR (Kollnig et al., 2021), greatly drawing into question its effectiveness. Therefore, the ultimate remedy to the data privacy challenges our age – if one exists

---

[29] Ibid.

- will require synergy between individual and collective, and regulatory and technological, solutions. It will require VcPAs and GDPRs working in harmony.

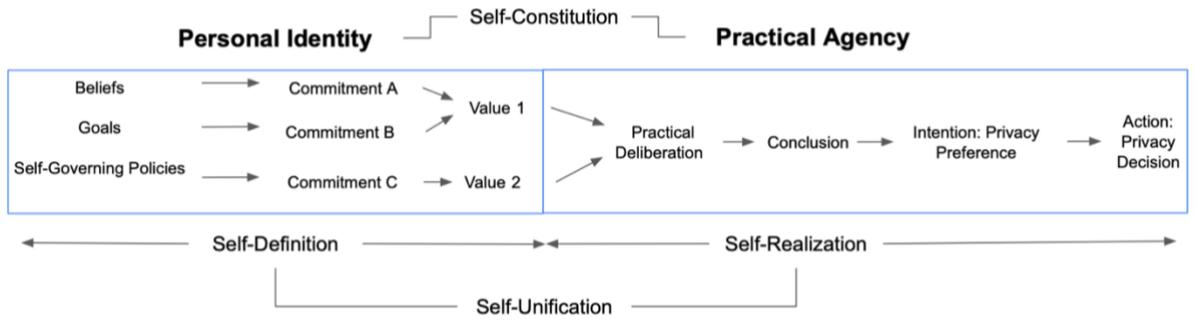

*Figure 1: Four-Dimensional Theory of Self-Governance (4DT) as it pertains to individual data privacy decision-making*

**Table 1:** Current personalized privacy assistant (PPA) evaluation using the Killmister's Four-Dimensional Theory of Autonomy (4DT) and suggested modifications for a value-centered privacy assistant (VcPA) for smartphone app selection

| Challenge | Relevant Dimensions of Autonomy | Do Current PPAs Address this Issue? | Suggested Modifications for a Value Centered, Smartphone Privacy Assistant (VcPA) |
|---|---|---|---|
| *Notice Fatigue* | Self-realization Self-unification Self-constitution | In part: Personalized and selective notifications help prevent *notice fatigue* and promote more value-consistent privacy choices through self-binding, but do not best prevent double-binds | Keep selective notices, but base profiles on user values rather than privacy preferences alone to help minimize double-binds |
| *Inertia Bias* | Self-realization Self-unification | No: the *inertia bias* makes it difficult for users to change from their initial privacy settings | An *exploratory process,* perhaps using exploratory notices, that does not encourage one download choice over another |
| *Lack of Controls* | Self-realization Self-unification | No: because of the *limited granularity problem*, a user's ability to realize their commitments and act on them is hindered | *Suggesting alternatives* to quickly link users to similar applications whose data collection practices better align with their values |